\documentclass[11pt, a4paper]{article}

\usepackage[T1,T2A]{fontenc}
\usepackage[cp1251]{inputenc}
\usepackage[english]{babel}

\usepackage{amsmath,euscript,mathrsfs}
\usepackage{amssymb,amsfonts,latexsym,mathtools}

\usepackage{graphicx,setspace}
\usepackage[usenames,dvipsnames,svgnames]{xcolor}
\usepackage[section]{placeins}
\graphicspath{{figures/}}

\usepackage[top=20mm,bottom=20mm,left=25mm,right=15mm]{geometry}
\usepackage{enumitem}

\usepackage[unicode]{hyperref}
\hypersetup{
	colorlinks=true,
	linkcolor={black!50!black},
	citecolor={blue!50!black},
	urlcolor={blue!50!black},
}

\DeclareMathOperator{\trace}{tr}
\DeclareMathOperator{\atanh}{atanh}

\allowdisplaybreaks[1]

\newcommand{\lsq}{\left [}
\newcommand{\rsq}{\right ]}

\newcommand{\vect}[1]{\underline{#1}}
\newcommand{\tens}[1]{\underline{\underline{#1}}}

\newcommand{\pdiff}[2]{\frac{\partial #1}{\partial #2}}

\begin{document}

\title{Longitudinal bulk strain solitons in a hyperelastic rod with quadratic and cubic nonlinearities\\
%Boussinesq and Korteweg -- de Vries Models with Third Order Nonlinearity for Longitudinal Strain Waves
{\large F.E.  Garbuzov$^1$, Y.M. Beltukov$^1$, K.R. Khusnutdinova${^{2, *}}$}\\[2ex]
{\small
$^1$ Ioffe Institute, 26 Polytekhnicheskaya, St. Petersburg 194021, Russia\\
$^2$ Department of Mathematical Sciences, Loughborough University, \\
Loughborough LE11 3TU, United Kingdom\\
$^*$ Corresponding author: K.Khusnutdinova@lboro.ac.uk\\[4ex]
{\it Dedicated to V.E. Zakharov on the occasion of his 80th birthday.}
}}
\date{}
\maketitle

\begin{abstract}
We study long nonlinear longitudinal bulk strain waves in a hyperelastic rod of circular cross section within the scope of the general weakly-nonlinear elasticity leading to a model with quadratic and cubic nonlinearities. We systematically derive the extended Boussinesq and Korteweg - de Vries - type equations and construct a family of approximate weakly-nonlinear soliton solutions with the help of near-identity transformations. These solutions are compared with the results of direct numerical simulations of the original nonlinear problem formulation, showing excellent agreement within the range of their asymptotic validity (waves of small amplitude) and extending their relevance beyond it (to the waves of moderate amplitude) as a very good initial guess. In particular, we were able to observe a stably propagating "table-top" soliton.
\end{abstract}

\noindent
%MSC: 35Q51, 35Q53

\noindent
%PACS: 62.30, 43.25

\noindent
%Keywords: Hyperelastic rod, Korteweg - de Vries - type equation, near-identity transformation, soliton

\section{Introduction}
Solitons have been a subject of a huge body of theoretical and experimental research in such areas as fluids and nonlinear optics,  largely because of the compact form of the governing equations  and availability of a large amount of experimental and observational data (see \cite{W, AS, NMPZ, N} and references therein). In contrast to that, the studies of solitary waves in solids is a relatively recent area of research, generally requiring greater efforts because of the complexity and great variability of the properties of solids reflected in their constitutive relations, as well as significant experimental challenges (for example, \cite{Maugin,Dai,M,HL,E1,P,E2} and references therein). Considerable progress has been made in the studies of bulk strain solitons in hyperelastic rods, starting with the works of G.A. Nariboli and A. Sedov \cite{NS} and L.A. Ostrovsky and A.M. Sutin \cite{OS}, and significantly advanced by A.M. Samsonov and his group (see \cite{S1, S2, SP, PV, S_book, P_book} and references therein). Theoretical studies were based on the Boussinesq and Korteweg-de Vries-type models developed within the scope of the weakly-nonlinear elasticity theory (Murnaghan's 5 constant model for elastic energy \cite{Murnaghan}), with differing degree of rigour. A systematic asymptotic analysis has been developed by H.-H. Dai and X. Fan \cite{DF} (although a systematic derivation of a Boussinesq-type equation was developed later, by F.E. Garbuzov et al. \cite{GKS}) and K.R. Khusnutdinova et al. \cite{KSZ}, within the scope of nonlinear elasticity and lattice modelling, respectively.  In \cite{GKS} the derivations within the scope of the general weakly-nonlinear elasticity theory have been simplified and generalised to include surface loading and longitudinal pre-stretch, resulting in the Boussinesq- and  forced Boussinesq-type models.
 The Boussinesq-type models have been used to study, in particular, the scattering of long longitudinal bulk strain solitary waves by delamination (see  \cite{KS, KT1, KT2}, and for related experiments see \cite{JAP2010, JAP2012}). 

In the present paper we aim to study elastic solitons of both small and moderate amplitude, and therefore we extend the derivation of nonlinear two-directional long wave models  for longitudinal waves   to hyperelastic materials described by the  9 constant model for the energy of the elastic deformation including cubic and quartic terms. We account for both geometrical and physical sources of nonlinearity and develop a systematic asymptotic analysis. The derivations are performed using symbolic computations with MATHEMATICA \cite{Mathematica}. We then derive a uni-directional extended Korteweg - de Vries (KdV) - type model and study its solitary wave solutions both analytically, with the help of near-identity transformations \cite{K, FL} (see also the review \cite{HK} and references therein) and direct numerical simulations of the original problem formulation.

\section{Problem formulation}

We consider a rod of circular cross section with the radius $R$ and use cylindrical coordinates $(x, r, \varphi)$ with the axial coordinate $x$, radial coordinate $r$ and angular coordinate $\varphi$. We use the Lagrangian description and denote the displacement vector by $\underline{U} = (U, V, W) $, where $ U $ is the axial displacement, $ V $ is the radial displacement and $ W $ is the torsion. 
\begin{figure}[h]
	\centering
	\includegraphics[width=0.3\textwidth]{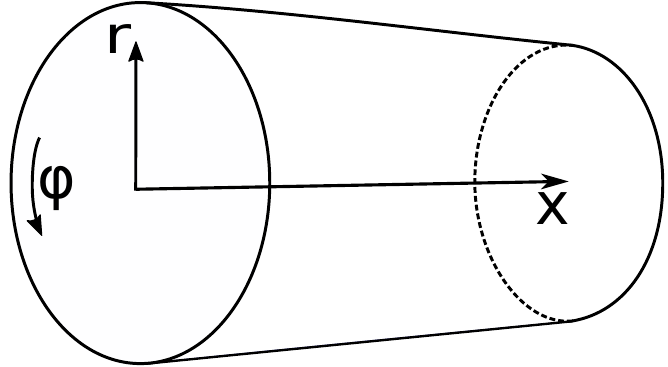}
	\caption{Rod of circular cross section.}
	\label{fig:rod}
\end{figure}

We use the fourth-order Landau-Lifshits constitutive relation \cite{LL} for the energy of the elastic deformation, which can be written as follows:
\begin{equation}\label{pot_landau}
\begin{split}
&\Pi = \frac{\lambda}{2}(\trace{\tens{\mathcal{E}}})^2 + \mu \trace{\tens{\mathcal{E}}^2} + \frac{A}{3}\trace\tens{\mathcal{E}}^3 + B\trace\tens{\mathcal{E}}\trace\tens{\mathcal{E}^2} + \frac{C}{3} (\trace\tens{\mathcal{E}})^3 \\
&\hspace{40mm}+ D\trace\tens{\mathcal{E}}\trace\tens{\mathcal{E}}^3 + F(\trace\tens{\mathcal{E}})^2\trace\tens{\mathcal{E}}^2 + G(\trace\tens{\mathcal{E}}^2)^2 + H(\trace\tens{\mathcal{E}})^4,
\end{split}
\end{equation}
where $\tens{\mathcal{E}} = (\nabla\vect{U}^T + \nabla\vect{U} + \nabla\vect{U}^T\cdot\nabla\vect{U})/2$ is the Cauchy-Green strain tensor.
This is equivalent to the Murnaghan 9~constant model~\cite{M}:
\begin{equation}\label{pot_murn}
\Pi = \frac{\lambda + 2\mu}{2}I_1^2 - 2\mu I_2 + \frac{l+2m}{3}I_1^3 - 2m I_1 I_2 + n I_3 + \nu_1 I_1^4 + \nu_2 I_1^2 I_2 + \nu_3 I_1 I_3 + \nu_4 I_2^2,
\end{equation}
where 
$I_1 = \trace{\tens{\mathcal{E}}}, \ I_2 = \lsq(\trace \tens{\mathcal{E}})^2 - \trace \tens{\mathcal{E}}^2\rsq/\,2,\ I_3 = \det \tens{\mathcal{E}}$,
 and 
 \begin{equation}
 l = B+C, \ m = A/2 + B, \ n = A, \ \nu_1 = D+F+G+H, \ \nu_2 = -(2F+3D+4G), \ \nu_3 = 3D, \ \nu_4 = 4G.
 \label{coefs}
 \end{equation}

We now consider an exact reduction of the full equations of motion describing solutions with no torsion, and where the longitudinal and transverse displacements $U$ and $V$ are independent of $ \varphi $:
\begin{equation}\label{assumptions}
U = U(x,r,t), \quad V = V(x,r,t), \quad W = 0.
\end{equation}
The equations of motion take the form
\begin{align}
\label{eq1_0}
&\rho  \frac{\partial^2 U(x,r,t)}{\partial t^2}-\frac{\partial P_{x x}}{\partial x}-\frac{\partial P_{xr}}{\partial r}-\frac{P_{xr}}{r} = 0,\\
\label{eq2_0}
&\rho  \frac{\partial^2 V(x,r,t)}{\partial t^2} - \frac{\partial P_{rx}}{\partial x}-\frac{\partial P_{rr}}{\partial r}-\frac{P_{rr} - P_{\varphi\varphi}}{r} = 0,
\end{align}
while the third equation is identically satisfied. Here, $P_{\alpha \beta}$ denotes components of the first Piola-Kirchhoff stress tensor 
\begin{equation} \label{piola}
\tens{P} = (\tens{I} + \nabla\vect{U}) \cdot \pdiff{\Pi}{\tens{\mathcal{E}}},
\end{equation}
where $\tens{I}$ is the identity tensor.

We assume that the rod is not subjected to any external loading, i.e. the stress has to vanish at the surface of the rod
\begin{align}
P_{rr} = P_{xr} = 0 \quad \mbox{at} \quad r = R. \label{bc}
\end{align}
Since the component $ P_{\varphi r} \equiv 0 $, the third boundary condition $P_{\varphi r} = 0$ at $r = R$ is identically satisfied.

We consider longitudinal waves in a symmetric rod, hence we add symmetry conditions which require the longitudinal displacement to be an even function of $r$ and the radial displacement to be an odd function of $r$ (e.g., \cite{GKS}).

\section{Extended Boussinesq-type equation}
We extend the approach developed in our previous paper~\cite{GKS}.
We look for a solution of the  problem in the form of power series expansions of the displacements in the radial coordinate:
\begin{eqnarray}
\label{u_series}
U(x,r,t) &=& U_0(x,t) + r^2 U_2(x,t) + r^4 U_4(x,t) + r^6 U_6(x,t) + \dots \, ,\\
\label{v_series}
V(x,r,t) &=& r V_1(x,t) + r^3 V_3(x,t) + r^5 V_5(x,t) + r^7 V_7(x,t) + \dots \, ,
\end{eqnarray}
which follow from the symmetry conditions.
We consider the waves of small amplitude and large length compared to the radius of the rod. Hence we non-dimensionalise the variables as follows:
\begin{equation} \label{scales1}
\tilde t = \frac{t}{L/c}, \quad \tilde x = \frac{x}{L}, \quad \tilde r = \frac{r}{\delta L}, \quad \tilde U = \frac{U}{\varepsilon L}, \quad \tilde V = \frac{V}{\varepsilon \delta L},
\end{equation}
which yields
$\displaystyle \tilde U_n =  \frac{L^n U_n}{\varepsilon L},  \ \tilde V_n =  \frac{L^n V_n}{\varepsilon L}$ for $n \ge 0$, 
assuming that $L$ is the characteristic wavelength, $c$ is the linear wave speed, $E$ is the Young modulus, $\varepsilon$ is the small amplitude parameter (characterising the longitudinal strain), and $\displaystyle \delta = \frac{R}{L} $ is the second small parameter (long wavelength parameter). Here, the tilde denotes dimensionless variables and tractions. In the following we will use expressions for the Young modulus and the Poisson ratio in terms of the Lame coefficients:
\begin{equation}\label{young_mod}
E = \frac{\mu(3\lambda + 2\mu)}{\lambda + \mu}, \quad \nu = \frac{\lambda}{2 (\lambda + \mu)}.
\end{equation}
Then, the expansions \eqref{u_series} and \eqref{v_series} take the form
\begin{align}
\label{u_series_scaled}
\widetilde U(\tilde x, \tilde r, \tilde t) &= \widetilde{U}_0(\tilde x, \tilde t) + \tilde{r}^2\widetilde{U}_2(\tilde x, \tilde t) + \tilde{r}^4\widetilde{U}_4(\tilde x, \tilde t) + O(\tilde r^6),\\
\label{v_series_scaled}
\widetilde V(\tilde x, \tilde r, \tilde t) &= \tilde{r} \widetilde{V}_1(\tilde x, \tilde t) + \tilde{r}^3\widetilde{V}_3(\tilde x, \tilde t) + \tilde{r}^5\widetilde{V_5}(\tilde x, \tilde t) + O(\tilde r^7).
\end{align}
In what follows we omit the tildes.

Substituting \eqref{u_series_scaled} and \eqref{v_series_scaled} into the equations of motion \eqref{eq1_0} and \eqref{eq2_0} we obtain
\begin{align} 
\label{eq1_1}
\begin{split}
&\rho c^2 U_{0tt} - (\lambda + 2\mu) U_{0xx} - 2(\lambda + \mu) V_{1x} - 4\mu U_2 + \Phi_{1,1}%(U_0, V_1, U_2) 
\varepsilon + \Phi_{1,2}%(U_0, V_1, U_2)
\varepsilon^2\\
&\quad+ \left[\rho c^2 U_{2tt} - (\lambda + 2\mu)U_{2xx} - 4(\lambda + \mu)V_{3x} - 16\mu U_4 + \Phi_{1,3}%(U_0, V_1, U_2, V_3, U_4)
\varepsilon\right] r^2\\ 
&\qquad + \left[\rho c^2 U_{4tt} - (\lambda + 2\mu)U_{4xx} - 6(\lambda + \mu)V_{5x} - 36\mu U_6\right] r^4 + O(\varepsilon^3, \varepsilon^2 r^2, \varepsilon r^4, r^6) = 0,
\end{split}\\
\label{eq2_1}
\begin{split}
&r \big( \rho c^2 V_{1tt} - \mu V_{1xx} - 2(\lambda + \mu)U_{2x} - 8(\lambda + 2\mu)V_3 + \Phi_{2,1}%(U_0, V_1, U_2, V_3)
\varepsilon + \Phi_{2,2}%(U_0, V_1, U_2, V_3)
\varepsilon^2 \\
&\quad- \left[\rho c^2 V_{3tt} - \mu V_{3xx} - 4(\lambda + \mu)U_{4x} - 24(\lambda + 2\mu)V_5 + \Phi_{2,3}%(U_0, V_1, U_2, V_3)
\varepsilon \right] r^2 \\
&\qquad- \left[\rho c^2 V_{3tt} - \mu V_{5xx} - 6(\lambda + \mu)U_{6x} - 48(\lambda + 2\mu)V_7 \right] r^4 + O(\varepsilon^3, \varepsilon^2 r^2, \varepsilon r^4, r^6)\big) = 0.
\end{split}
\end{align}
Here, the subscripts $x$ and $t$ denote partial derivatives and $\Phi_{i,1}$, $\Phi_{i,2}$, $\Phi_{i,3}$ denote all nonlinear terms with the coefficients $\varepsilon$, $\varepsilon^2$ and $\varepsilon r^2$, respectively.
The functions $ U_2 $, $ V_3 $, $ U_4 $, $V_5$, $U_6$ can be obtained using the power series expansions in  $\varepsilon$:
$$ U_2 = U_2^{(0)} + \varepsilon U_2^{(1)} + \varepsilon^2 U_2^{(2)} + \dots.$$
Equating to zero the coefficients at different powers of $\varepsilon$ and $r$ in \eqref{eq1_1} and \eqref{eq2_1} results in
\begin{align}
\label{U2}
U_2 &= \frac{1}{4\mu} \left[ \rho c^2 U_{0tt} - (\lambda + 2\mu) U_{0xx} - 2(\lambda + \mu) V_{1x} \right] + \varepsilon U_2^{(1)}(x,t) + \varepsilon^2 U_2^{(2)}(x,t) + O(\varepsilon^3),\\
\label{V3}
V_3 &= \frac{1}{8(\lambda + 2\mu)} \left[ \rho c^2 V_{1tt} - 2(\lambda + \mu) U_{2x} - \mu V_{1xx} \right] + \varepsilon V_3^{(1)}(x,t) + \varepsilon^2 V_3^{(2)}(x,t) + O(\varepsilon^3),\\
\label{U4}
U_4 &= \frac{1}{16\mu}\left[\rho c^2 U_{2tt} - (\lambda + 2\mu) U_{2xx} - 4(\lambda + \mu) V_{3x}\right] + \varepsilon U_4^{(1)}(x,t) + O(\varepsilon^2),\\
V_5 &= \frac{1}{24(\lambda + 2\mu)} \left(\rho c^2 V_{3tt} - 4(\lambda+\mu)U_{4x} - \mu V_{3xx}\right)  + \varepsilon V_5^{(1)}(x,t) + O(\varepsilon^2),\\
U_6 &= \frac{1}{36\mu}\left[\rho c^2 U_{4tt} - (\lambda + 2\mu) U_{4xx} - 6(\lambda + \mu) V_{5x}\right] + O(\varepsilon).
\end{align}
The expressions for the functions $U_2^{(1)}$, $U_2^{(2)}$, $V_3^{(1)}$, $V_3^{(2)}$, $U_4^{(1)}$, $V_5^{(1)}$ are cumbersome and are not shown here.
Next, substituting the functions $ U_2 $, $ V_3 $, $ U_4 $, $V_5$, $U_6$ into the boundary conditions \eqref{bc} we obtain the equations
\begin{align} 
\label{bc_rr_subst}
\begin{split}
&2 (\lambda + \mu) V_1 + \lambda U_{0x} + \varepsilon \Psi_{1,1} + \varepsilon^2 \Psi_{1,2}  + \delta^2 \bigg[ d_1 U_{0xxx} + \rho c^2 d_2 U_{0xtt} + \rho c^2 d_3 V_{1tt} + d_4 V_{1xx}\bigg] \\
&\qquad + \delta^4\Big[\left(d_5 V_{1xx} + \rho c^2 d_6 V_{1tt}\right)_{xx} + \rho^2 c^4\left(d_7 U_{0x} + d_8 V_{1}\right)_{tttt} + \left(d_9 U_{0xx} + \rho c^2 d_{10} U_{0tt}\right)_{xxx}\Big]\\
&\hspace{93mm} + \varepsilon\delta^2\Psi_{1,3} + O(\varepsilon^3, \varepsilon^2 \delta^2, \varepsilon \delta^4, \delta^6) =  0,
\end{split}\\
\label{bc_rx_subst}
\begin{split}
&\rho  c^2 U_{0tt} -2 \lambda  V_{1x}-(\lambda +2 \mu ) U_{0xx} + \varepsilon \Psi_{2,1} + \varepsilon^2 \Psi_{2,2} + \varepsilon\delta^2\Psi_{2,3}\\
&\hspace{12mm}+ \delta^2\Big[e_1 U_{0xxxx} + \rho^2 c^4 e_2 U_{0tttt} + \rho c^2 e_3 U_{0xxtt} + e_4 V_{1xxx} + \rho  c^2 e_5 V_{1xtt} \Big]\\
&\hspace{32mm}+ \delta^4 \Big[ \left(e_6 V_{1xx} + \rho c^2 e_7 V_{1tt}\right)_{xxx} + \rho^2c^4\left(e_8 V_{1x} + e_9 U_{0xx} + \rho c^2 e_{10} U_{0tt}\right)_{tttt} \\
&\hspace{64mm}+ \left(e_{11} U_{0xx} - \rho c^2 U_{0tt}\right)_{xxxx} \Big] + O(\varepsilon^3, \varepsilon^2 \delta^2, \varepsilon \delta^4, \delta^6)
= 0.
\end{split}
\end{align}
Here the coefficients $d_i$, $e_i$ depend on the Lame elastic moduli, and $\Psi_{i,j}$ denote nonlinear terms.
Elimination of the function $V_1$ from the equations \eqref{bc_rr_subst} and \eqref{bc_rx_subst} can be done by expanding it into the power series in $\varepsilon$ and $\delta^2$. Unknown terms in this expansion can be found by equating to zero the coefficients of $\varepsilon$, $\delta^2$, $\varepsilon^2$, $\delta^4$ and $\varepsilon\delta^2$ in \eqref{bc_rr_subst}:
\begin{equation} \label{v1_asympt}
\begin{split}
V_1(x, t) =& - \frac{\lambda}{2(\lambda + \mu)} U_{0x} + \varepsilon f(x,t) + \delta^2 g(x,t) + \varepsilon^2 \tilde f(x,t) + \delta^4 \tilde g(x,t) + \varepsilon\delta^2 \tilde h(x,t)\\
& + O(\varepsilon^3, \varepsilon^2 \delta^2, \varepsilon \delta^4, \delta^6).
\end{split}
\end{equation}
Here we do not show the expressions for the functions $f$, $g$, $\tilde f$, $\tilde g$, $\tilde h$, for brevity. Then, the substitution of $V_1$ into \eqref{bc_rx_subst} results in the following equation for $U_0$:
\begin{equation}\label{eq_u0_fin}
%\nonumber
\begin{split}
&U_{0tt} - U_{0xx} + \varepsilon \frac{\beta_1}{E}\left(U_{0x}^2\right)_x + \varepsilon^2 \frac{\beta_2}{E^2}\left(U_{0x}^3 \right)_x + \delta^2 \left[\alpha_1 U_{0tttt} + \alpha_2 U_{0xxtt} + \alpha_3 U_{0xxxx} + \varepsilon \widetilde J_0(U_0)\right]\\
&\hspace{20mm}+ \delta^4\left(\alpha_4 U_{0tttttt} + \alpha_5 U_{0xxtttt} + \alpha_6 U_{0xxxxtt} + \alpha_7 U_{0xxxxxx}\right) + O(\varepsilon^3, \varepsilon^2 \delta^2, \varepsilon \delta^4, \delta^6) = 0,
\end{split}
\end{equation}
where the coefficients $\alpha_i$, $\beta_i$ and the nonlinear function $\widetilde J_0$ are given in Appendix~A. Then, assuming the balance between the nonlinear and dispersive terms $\varepsilon\sim\delta^2$, and truncating this equation, we obtain an extended Boussinesq-type equation:
\begin{equation}\label{eq_u0_fin_trunc}
\begin{split}
&U_{0tt} - U_{0xx} + \varepsilon \Big[ \frac{\beta_1}{E}\left(U_{0x}^2\right)_x + \alpha_1 U_{0tttt} + \alpha_2 U_{0xxtt} + \alpha_3 U_{0xxxx}\Big] \\
&\quad + \varepsilon^2 \Big[\frac{\beta_2}{E^2}\left(U_{0x}^3\right)_x + \alpha_4 U_{0tttttt} + \alpha_5 U_{0xxtttt} + \alpha_6 U_{0xxxxtt} + \alpha_7 U_{0xxxxxx} + \widetilde J_0(U_0)\Big] = 0.
\end{split}
\end{equation}
The equation (\ref{eq_u0_fin_trunc}) can be rewritten in a simpler asymptotically equivalent form
\begin{equation}\label{eq_u0_fin_trunc_reg}
\begin{split}
&U_{0tt} - U_{0xx} + \varepsilon \Big[ \frac{\beta_1}{E}\left(U_{0x}^2\right)_x + q_1 U_{0xxtt}\Big] + \varepsilon^2 \Big[\frac{\beta_2}{E^2}\left(U_{0x}^3\right)_x + \frac{q_2 - 2\gamma_1 - 4\gamma_4}{E} U_{0xx}U_{0xxx} \\
&\hspace{55mm} + \frac{2\gamma_1 + 4\gamma_4}{E} U_{0xt}U_{0xxt} + \frac{q_3}{E} U_{0x}U_{0xxxx} + q_4 U_{0xxxxtt}\Big] = 0,
\end{split}
\end{equation}
where 
\begin{eqnarray}
&&q_1 = \alpha_1 + \alpha_2 + \alpha_3, \ q_2 = 3\gamma_1 + 2\gamma_2 + 3\gamma_3 + 6\gamma_4 + \gamma_5 + 6\gamma_6 + 2\gamma_7, \nonumber  \\
&& q_3 = \gamma_1 + \gamma_3 + 2\gamma_4 + \gamma_5 + 2\gamma_6, \ q_4 = \alpha_4 + \alpha_5 + \alpha_6 + \alpha_7. \label{qs}
\end{eqnarray}
The dimensional form of  the equation \eqref{eq_u0_fin_trunc_reg} is given by
\begin{equation}\label{eq_u0_fin_trunc_reg_dim}
\begin{split}
&U_{0tt} - c^2 U_{0xx} + \frac{\beta_1}{\rho}\left(U_{0x}^2\right)_x + q_1 R^2 U_{0xxtt} + \frac{\beta_2}{E\rho}\left(U_{0x}^3\right)_x + \frac{(q_2 - 2\gamma_1 - 4\gamma_4)R^2}{\rho} U_{0xx}U_{0xxx} \\
&\hspace{45mm} + \frac{(2\gamma_1 + 4\gamma_4)R^2}{E} U_{0xt}U_{0xxt} + \frac{q_3 R^2}{\rho} U_{0x}U_{0xxxx} + q_4 R^4 U_{0xxxxtt} = 0.
\end{split}
\end{equation}
A particular case of this equation has been considered in \cite{CLS, SCL1, SCL2}.

We note that, unlike \cite{PM}, the equation obtained using a systematic asymptotic procedure contains non only the additional cubic nonlinear term, but also several other nonlinear and dispersive terms. The expression for the coefficient in front of the cubic nonlinearity in \cite{PM} has a typo. The correct formula can be found in Appendix A (given in terms of the Landau moduli, relations between the Landau and Murnaghan moduli can be found in (\ref{coefs})).

\section{Extended Korteweg - de Vries - type equation and solitons}
%First, let us consider the case of free lateral surface: $P=T=0$. 
We introduce the characteristic variables $\xi = x - t, \ \eta = x + t, \ \tau = \varepsilon t$ and expand the unknown function $U_0$ into the power series in $\varepsilon$. In order to derive a uni-directional model we look for a solution in the form
%assume that the first term in this series is independent of $\eta$, which leads to independence of $\eta$ of all subsequent terms:
% introduce the new variables and assume that unknown displacement $U_0$ is dependant on them only:
\begin{equation}\label{u0_expansion}
U_0(x,t) = U_0^{(0)}(\xi, \tau) + \varepsilon U_0^{(1)}(\xi, \tau) + \varepsilon^2 U_0^{(2)}(\xi, \tau) + \dots,
\end{equation}
which can be justified by allowing the dependence of the higher-order terms on both characteristic variables and requiring the absence of secular terms in the asymptotic expansion (e.g. \cite{N}).
Substitution of $U_0$ into the equation \eqref{eq_u0_fin_trunc} (one could also use (\ref{eq_u0_fin_trunc_reg})) yields 
\begin{equation}\label{eq_kdv_proto}
\begin{split}
&\left(U_{0\tau}^{(0)} - \frac{\beta_1}{2E} U_{0\xi}^{(0)2} - \frac{q_1}{2} U_{0\xi\xi\xi}^{(0)}\right)_\xi + \varepsilon \bigg[\left(U_{0\tau}^{(1)} - \frac{\beta_1}{2E} U_{0\xi}^{(0)} U_{0\xi}^{(1)} - \frac{q_1}{2} U_{0\xi\xi\xi}^{(1)}\right)_\xi + (2\alpha_1 + \alpha_2) U_{0\xi\xi\xi\tau}^{(0)} \\
&\qquad- \frac12 U_{0\tau\tau}^{(0)} - \frac{\beta_2}{2E^2}\left(U_{0\xi}^{(0)3}\right)_\xi - \frac{q_2}{2E} U_{0\xi\xi}^{(0)}U_{0\xi\xi\xi} - \frac{q_3}{2E} U_{0\xi}^{(0)}U_{0\xi\xi\xi\xi}^{(0)} - \frac{q_4}{2} U_{0\xi\xi\xi\xi\xi\xi}^{(0)}\bigg] + O(\varepsilon^2) = 0,
\end{split}
\end{equation}
where the coefficients $q_{1,2,3,4}$ are given in \eqref{qs}.
%where $q_1 = \alpha_1 + \alpha_2 + \alpha_3$, $q_2 = 3\gamma_1 + 2\gamma_2 + 3\gamma_3 + 6\gamma_4 + \gamma_5 + 6\gamma_6 + 2\gamma_7$, $q_3 = \gamma_1 + \gamma_3 + 2\gamma_4 + \gamma_5 + 2\gamma_6$, $q_4 = \alpha_4 + \alpha_5 + \alpha_6 + \alpha_7$. 
The $\tau$-derivatives in the $O(\varepsilon)$ terms in \eqref{eq_kdv_proto} can be eliminated using the asymptotic relation
$ U_{0\tau}^{(0)} = \frac{\beta_1}{2E} U_{0\xi}^{(0)2} + \frac{q_1}{2} U_{0\xi\xi\xi}^{(0)} + O(\varepsilon). $
Then, introducing the new function $u = U_{0\xi}^{(0)} + \varepsilon U_{0\xi}^{(1)}$ we obtain
\begin{equation}\label{eq_ekdv}
\begin{split}
&u_\tau - \frac{\beta_1}{2E} \left(u^{2}\right)_\xi - \frac{q_1}{2} u_{\xi\xi\xi} - \varepsilon \bigg[\frac{3\beta_2 + \beta_1^2}{6E^2}\left(u^{3}\right)_\xi + \frac{2q_2 + 3(q_1 - 4(2\alpha_1 + \alpha_2))\beta_1}{4E} u_{\xi}u_{\xi\xi}\\
&\qquad + \frac{q_3 + (q_1 - 2(2\alpha_1+\alpha_2))\beta_1}{2E} u u_{\xi\xi\xi} + \left(\frac{q_4}{2} - \frac{q_1(2\alpha_1 + \alpha_2)}{2} + \frac{q_1^2}{8}\right) u_{\xi\xi\xi\xi\xi}\bigg] + O(\varepsilon^2) = 0.
\end{split}
\end{equation}
 The dimensional form of the equation \eqref{eq_ekdv} is given by
\begin{equation}\label{eq_ekdv_dim}
\begin{split}
&\frac{1}{c} u_t - \frac{\beta_1}{2E} \left(u^{2}\right)_\xi - \frac{q_1 R^2}{2} u_{\xi\xi\xi} - \frac{3\beta_2 + \beta_1^2}{6E^2}\left(u^{3}\right)_\xi - \frac{[2q_2 + 3(q_1 - 4(2\alpha_1 + \alpha_2))\beta_1]R^2}{4E} u_{\xi}u_{\xi\xi}\\
&\qquad - \frac{[q_3 + (q_1 - 2(2\alpha_1+\alpha_2))\beta_1] R^2}{2E} u u_{\xi\xi\xi} - \left(\frac{q_4}{2} - \frac{q_1(2\alpha_1 + \alpha_2)}{2} + \frac{q_1^2}{8}\right)R^4\, u_{\xi\xi\xi\xi\xi} = 0,
\end{split}
\end{equation}
where the dimensional $\xi = x - c t$ and $c^2 = E/\rho$.
Note that since $U$ is a longitudinal displacement the function $u$ can be treated as a longitudinal strain. The equation (\ref{eq_ekdv}) has been derived and studied mainly in the context of waves in fluids (see \cite{B, LB, KB, LY, GPP, GKKS, KRI, KST} and references therein), and is often referred to as an extended Korteweg - de Vries (eKdV) equation. To the best of our knowledge, this is the first derivation of this equation in the context of waves in solids. Some other extended models have been obtained in \cite{D, DH}. 

Studying solitary wave solutions of the derived equation directly is a complicated task, and therefore here we aim to reduce the eKdV equation \eqref{eq_ekdv} to the Gardner equation using direct and inverse near-identity transformations of the form 
\begin{align}
&\hat u = u + \varepsilon\big(a_1 u_{\xi\xi} + a_2\xi u_\tau + a_3 u_\xi \int_{\xi_0}^{\xi}u d\xi\big), \\
\label{near_ident}
&u = \hat u - \varepsilon\big(a_1 \hat u_{\xi\xi} + a_2\xi \hat u_\tau + a_3 \hat u_\xi \int_{\xi_0}^{\xi}\hat u d\xi\big),
\end{align}
up to $O(\varepsilon^2)$ corrections. We note that the general near-identity transformations discussed in~\cite{K, FL, HK} contain also the $\varepsilon a_4 u^2$ term which we do not use here since we wish to retain both the quadratic and cubic nonlinearities in the equation \eqref{eq_ekdv}. It is well-known from the studies in fluids that the Gardner equation has a rich family of solitons reducing to KdV solitons in the case of small amplitude \cite{GPP}. We would like to use these known solutions and to compare the related analytical solutions of our derived model with the results of direct numerical simulations of the original problem formulation. Various near-identity transformations have been used to study nonlinear waves in two- and three-layered fluids (e.g., \cite{MS2, MS3, PS, RKK_book}). We note that in the context of solids there are 9 free parameters (constants characterising elastic properties of various materials), and generally there is more freedom in the choice of the coefficients of the equation (\ref{eq_ekdv}) than in the known fluid contexts. 

The appropriate choice of the coefficients allows us to eliminate all higher-order dispersive terms from \eqref{eq_ekdv}:
\begin{equation}\nonumber
\begin{split}
&a_1 = \frac{1}{12} \left[\frac{10 q_4}{q_1} + q_1 + 2(2\alpha_1 + \alpha_2) + \frac{3(q_3 - q_2)}{\beta_1}\right]\hspace{-1.2mm}, \\ 
&a_2 = \dfrac{4q_4 - 4q_1(2\alpha_1 + \alpha_2) + q_1^2}{6q_1^2}, \quad
a_3 = \frac{\beta_1 q_1 (4\alpha_1 + 2\alpha_2 + q_1) + 3q_1 q_3 - 8\beta_1 q_4}{9 E q_1^2}.
\end{split}
\end{equation} 
The resulting Gardner equation takes the form:
\begin{equation}\label{eq_gardner}
\hat u_\tau - \frac{\beta_1}{2E} \left(\hat u^{2}\right)_\xi - \frac{q_1}{2} \hat u_{\xi\xi\xi} - \varepsilon \frac{\hat\beta_2}{2E^2}\left(\hat u^{3}\right)_\xi + O(\varepsilon^2) = 0,
\end{equation}
where $\hat\beta_2 = \beta_2 + (\beta_1^2(1 - 2a_2) - a_3 E \beta_1)/3$. 
The equation \eqref{eq_gardner} has a family of solitary wave solutions parametrised by the amplitude parameter $M$ (e.g. \cite{GPP}):
\begin{equation}\label{sol_gardner}
\hat u(\xi, t) = \frac{M}{1 + N\cosh K\theta}, \quad N = \sqrt{1 + \frac{3\varepsilon\hat\beta_2 M}{2\beta_1 E}}, \quad K = \sqrt{\frac{\beta_1 M}{3Eq_1}}, \quad v = -\frac{\beta_1 M}{6E},
\end{equation}
where $\theta = \xi - v t$.
We use this solution to create an asymptotic solution of  the originally derived extended KdV equation \eqref{eq_ekdv} with the accuracy up to $O(\varepsilon^2)$ terms using the inverse near-identity transformation~\eqref{near_ident}:
%\begin{equation}\label{sol_ekdv}
%u = \hat u - \left(R^2 a_1 \hat u_{\xi\xi} + \frac{a_2}{c}\xi \hat u_t + a_3 \hat u_\xi \int_{\xi_0}^{\xi}\hat u d\xi\right)
%\end{equation}
\begin{equation}\label{sol_ekdv}
\begin{split}
u(\xi, t) &= \frac{M}{1 + N\cosh K\theta}\Bigg[1 - \frac{\varepsilon a_1 N K^2 \left[N(\cosh 2K\theta - 3) - 2\cosh K\theta\right]}{2(1 + N\cosh K\theta)^2} \\
&\hspace{30mm}+ \frac{\varepsilon N K \sinh K\theta}{1 + N\cosh K\theta}\left(-a_2\xi v + \frac{2a_3 M \atanh \left(\sqrt\frac{1-N}{1+N} \tanh\frac K 2 \theta \right)}{\sqrt{1 - N^2}}\right)\Bigg].
\end{split}
\end{equation}
We note that the term which contains $\xi$ explicitly is not secular, because it is of the same order as $\xi/\cosh K\xi$, which decays to $0$ as $\xi\to\infty$ (this term can be removed by a phase shift). The dimensional form of the solutions \eqref{sol_gardner} and \eqref{sol_ekdv}, respectively, is as follows
\begin{align}
\label{sol_gardner_dim}
\hat u(\xi, t) &= \frac{M}{1 + N\cosh K\theta}, \quad N = \sqrt{1 + \frac{3\hat\beta_2 M}{2\beta_1 E}}, \quad K = \sqrt{\frac{\beta_1 M}{3Eq_1R^2}}, \quad v = -\frac{\beta_1 M c}{E};\\
\label{sol_ekdv_dim}
\begin{split}
u(\xi, t) &= \frac{M}{1 + N\cosh K\theta}\Bigg[1 - \frac{a_1 N K^2 R^2 \left[N(\cosh 2K\theta - 3) - 2\cosh K\theta\right]}{2(1 + N\cosh K\theta)^2} \\
&\hspace{30mm}+ \frac{N K \sinh K\theta}{1 + N\cosh K\theta}\left(-\frac{a_2 R^2 \xi v}{c} + \frac{2a_3 M \atanh \left(\sqrt\frac{1-N}{1+N} \tanh\frac K 2 \theta \right)}{\sqrt{1 - N^2}}\right)\Bigg].
\end{split}
\end{align}

It is now interesting to compare the performance of the simple Gardner soliton~\eqref{sol_gardner} and the formula \eqref{sol_ekdv} for  the solution of the original extended KdV equation both within the range of  its formal asymptotic validity (i.e. waves of small amplitude), as well as the case when cubic and quadratic nonlinear terms become comparable ($\varepsilon\beta_2 \sim \beta_1$, waves of moderate amplitude),  and also  to check whether the formula \eqref{sol_ekdv} is a better approximation to the numerical solution of the original problem than the simple Gardner soliton~\eqref{sol_gardner}. We note that, strictly speaking,  the case of waves of moderate amplitude is beyond the range of validity of the asymptotic expansion, however it is interesting to test whether the asymptotic formula can still be useful in one way or another.
We note that weakly-nonlinear solutions have been compared with the results of direct numerical simulations of the original problem formulations in several settings relevant to the oceanic studies (see \cite{RKK_book, PPL, L_et_al, M_et_al, M_et_al1, T_et_al} and references therein). To the best of our knowledge there were no comparisons for solids.

\begin{figure}[h]
	\centering
	\includegraphics[width=0.7\linewidth]{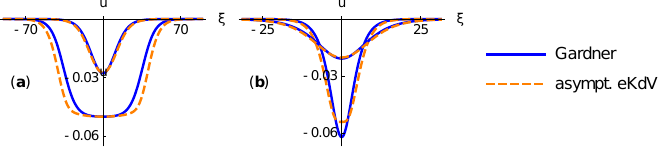}
	\caption{Comparison of the Gardner solitons \eqref{sol_gardner_dim} and the eKdV asymptotic solutions \eqref{sol_ekdv_dim} in dimensional variables for two materials: (a) Material 1, $M = -0.04$ (small amplitude) and $M = -0.049982$ (moderate amplitude); (b) Material 2, $M = -0.05$ (small amplitude) and $M = -0.2$ (moderate amplitude).}
	\label{fig:g_ekdv_sol}
\end{figure}

We consider two hyperelastic materials with the elastic moduli given in Table~\ref{tab:moduli}. Here for brevity we study only solitons of negative polarity (solitons of compression), hence we choose the moduli A, B and C so that  the coefficient of quadratic nonlinearity is negative. The Material 1 has a negative coefficient of cubic nonlinearity (GE$-$ case), therefore the corresponding family of small-amplitude solitons contains ``table-top" solitons. This coefficient for the Material 2 is positive (GE$+$ case). We note that $q_1 = -\nu^2/2 < 0$, hence the dispersive coefficient in the Gardner equation \eqref{eq_gardner} is always positive. Examples of solitons  in both materials are given in Figure \ref{fig:g_ekdv_sol}.

\vspace{-3mm}
\begin{table}[h]
	%\captionsetup{justification=raggedleft,singlelinecheck=false}
	\caption{Elastic moduli.}
	%\vspace{-6mm}
	\begin{center}
		\begin{tabular}{|c|c|c|c|c|c|c|c|c|c|c|c|c|}
			\hline
			&\rule[-1ex]{0pt}{3ex} Young's m. & Poisson's & \multicolumn{7}{|c|} {Landau moduli, GPa} & Density & \multicolumn{2}{|c|}{Coefficients}\\
			\cline{4-10}\cline{12-13}
			&\rule[-1ex]{0pt}{3ex} $E$, GPa & ratio, $\nu$ & $A$ & $B$ & $C$ & $D$ & $F$ & $G$ & $H$ & $\rho$, kg/m\textsuperscript{3} & $\beta_1/E$ & $\hat\beta_2/E^2$ \\
			\hline
			Material 1&\rule[-1ex]{0pt}{3ex} $5$ & $0.34$ & \multicolumn{3}{|c|}{$-5.85$} & \multicolumn{4}{|c|}{$-2.93$} & $1000$ & 1 & $13.3$ \\
			\hline
			Material 2&\rule[-1ex]{0pt}{3ex} $5$ & $0.34$ & \multicolumn{3}{|c|}{$-5.85$} & \multicolumn{4}{|c|}{$14.18$} & $1000$ & 1 & $-13.3$ \\
			\hline
		\end{tabular}
	\end{center}
	\label{tab:moduli}
\end{table}
\vspace{-4mm}

\section{Numerical simulations}

In order to compare the derived asymptotic solutions with the results of direct numerical simulations of the original problem formulation \eqref{eq1_0} -- \eqref{bc} we use a multidomain pseudospectral method  \cite{CHQZ3}. A~set of Legendre polynomials in both $x$ and $r$ variables are used for the spatial discretization of the problem:
$ \vect{U}(x,r,t) = \sum_{n,m} \vect{\widehat{U}}_{nm}(t) \Phi_{n}(x) \Psi_{m}(r),$
where $\vect{U}$ is the displacement vector. 
%The multidomain method allows us  to compute the solution relatively quickly on a fine mesh, which consists of 20 domains with 30 points in $x$ and 5 points in $r$ %each (the rod we consider is thin compared to the wavelength, hence we do not need a lot of points in $r$ coordinate).
The multidomain method allows us  to compute the solution relatively quickly on a fine mesh with 600-650 points in $x$, split into 20-25 domains,  and 5 points in $r$ (the rod we consider is thin compared to the wavelength, hence we do not need a large number of points in the $r$ coordinate).

%The multidomain method allows us  to compute the solution relatively quickly on a fine mesh with 650 points in $x$ and 5 points in $r$ (the rod we consider is thin compared to the wavelength, hence we do not need a lot of points in $r$ coordinate).

\begin{figure}[h]
	\centering
	\includegraphics[width=.8\linewidth]{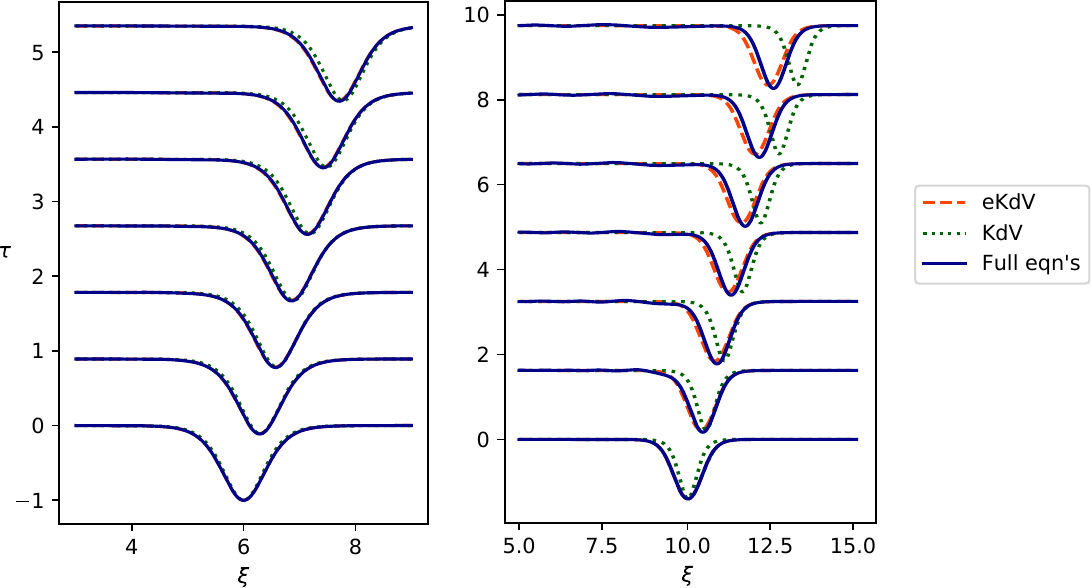}
	\caption{Evolution of the initial soliton \eqref{sol_ekdv} in the Material 1 (GE$-$ case). Nondimensional variables $\xi$, $\tau$. Amplitude of the initial nondimensional soliton equals 1; $\varepsilon = 0.005$ in the left plot, $\varepsilon = 0.027$ in the right plot (small amplitude).}
	\label{fig:sol_evol1_GE-}
\end{figure}

\begin{figure}[h]
	\centering
	\includegraphics[width=.8\linewidth]{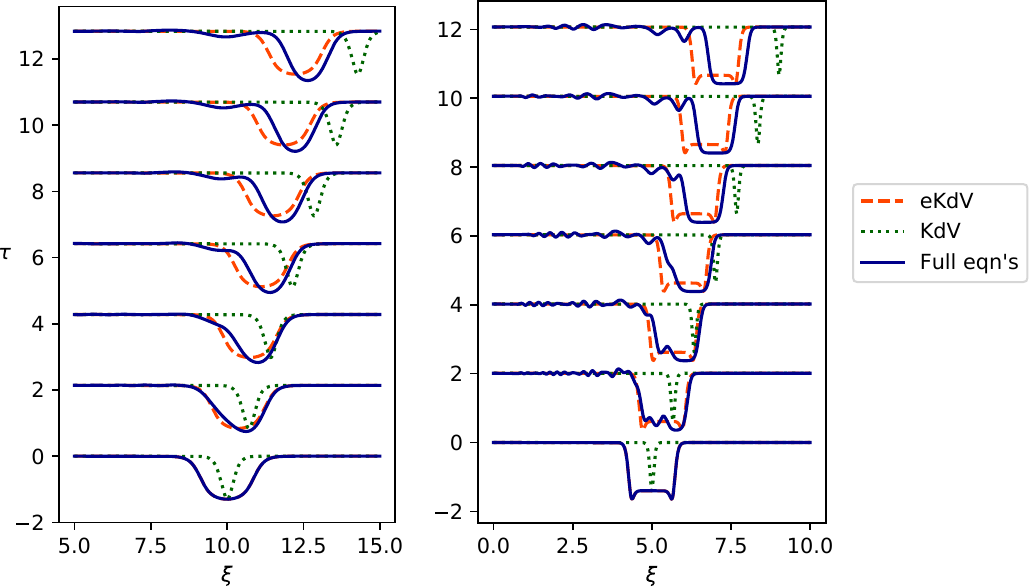}
	\caption{Evolution of the initial soliton \eqref{sol_ekdv} in the Material 1 (GE$-$ case). Nondimensional variables $\xi$, $\tau$. Amplitude of the initial nondimensional soliton equals 1; $\varepsilon = 0.0499$ in the left plot, $\varepsilon = 0.0499825675$ in the right plot (moderate amplitude). }
	\label{fig:sol_evol2_GE-}
\end{figure}

\begin{figure}[h]
	\centering
	\includegraphics[width=.8\linewidth]{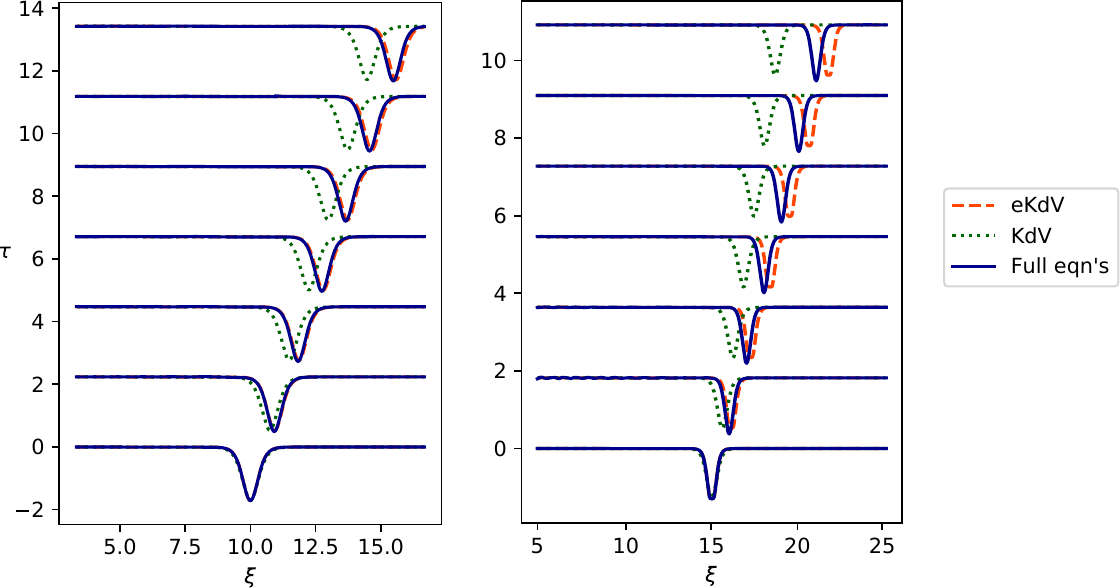}
	\caption{Evolution of the initial soliton \eqref{sol_ekdv} in Material 2 (GE+ case). Nondimensional variables $\xi$, $\tau$. Amplitude of the initial nondimensional soliton equals 1; $\varepsilon = 0.02$ in the left plot (small amplitude),  $\varepsilon = 0.05$ in the right plot (moderate amplitude).}
	\label{fig:sol_evol_GE+}
\end{figure}

\begin{figure}[h]
	\centering
	\includegraphics[width=.8\linewidth]{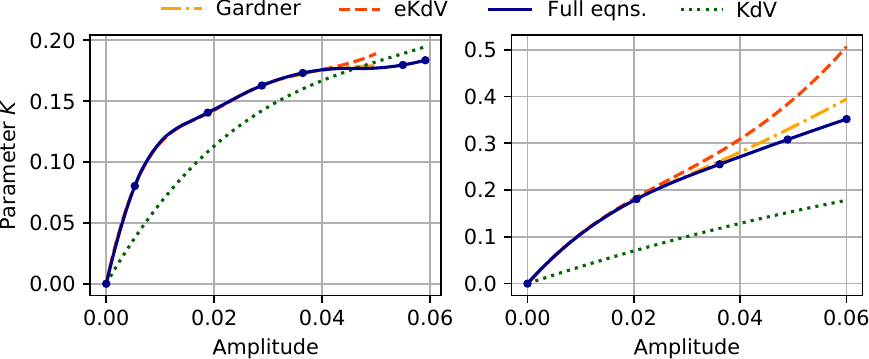}
	
	\vspace*{5mm}
	\includegraphics[width=.8\linewidth]{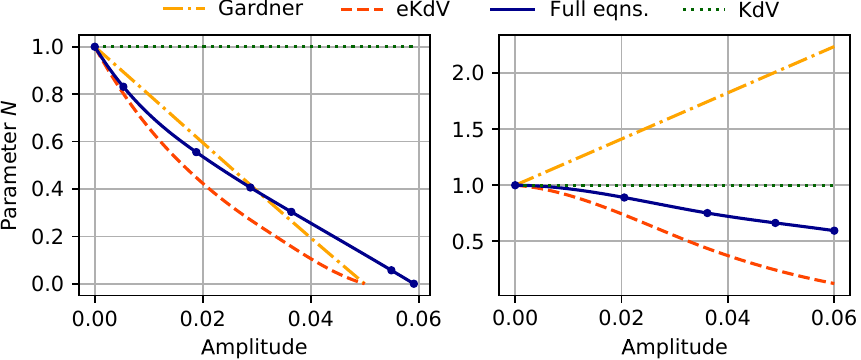}
	
	\vspace*{5mm}
	\includegraphics[width=.8\linewidth]{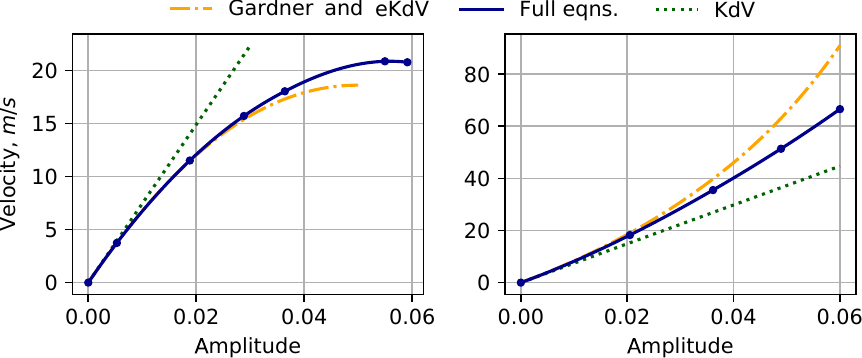}
	\caption{Relations between the soliton parameters. The left / right  plots correspond to the GE-- / GE+ case. All solitons are approximated by the function $\dfrac{M}{1 + N \cosh K(x - v t)}$.  The dots on the solid curve indicate results obtained from a set of numerical simulations. The cubic spline interpolation was used to draw the numerical curves.}
	\label{fig:params_rel}
\end{figure}

In Figures \ref{fig:sol_evol1_GE-}, \ref{fig:sol_evol2_GE-}, \ref{fig:sol_evol_GE+} the numerical results for the original problem formulation are obtained using the initial condition in the form of the eKdV asymptotic solution \eqref{sol_ekdv}. The initial soliton and the KdV soliton are plotted for comparison. The data for the soliton's velocity and shape is summarised in Figure~\ref{fig:params_rel}. From these comparisons we can see that overall the solution (\ref{sol_ekdv}) performs better than the solution of the KdV or Gardner equation, although all three solutions work very well for the case of waves of small amplitude, and the Gardner soliton is also a physically relevant approximation in the case of waves of moderate amplitude. 
Although the initial condition in the form of \eqref{sol_ekdv} has "horns" in the case of waves of moderate amplitude (Fig. 4), which seems to be physically irrelevant, this initial condition allowed us to very quickly generate and observe a ``table-top" soliton of the original problem formulation. We noticed that in this case more energy of the initial wave was transferred to this ``table-top" soliton and less energy was radiated away compared to the  case of the Gardner initial soliton \eqref{sol_gardner}.
From our experiments we can conclude that, at moderate amplitude, the constructed weakly-nonlinear solution can be used at least as a very good initial guess in order to generate moderately-nonlinear soliton solutions.

\section{Conclusions}

In this paper we derived the extended Boussinesq and Korteweg - de Vries  equations describing long nonlinear longitudinal bulk strain waves in generic weakly-nonlinear hyperelastic materials with the accuracy up to an including the cubic terms in the equations. The extended Korteweg - de Vries equation was then reduced to the Gardner equation with the help of a near-identity transformation in order to make use of the known family of soliton solutions of this equation. The inverse near-identity transformation was used to obtain the solution of the derived extended Korteweg - de Vries equation. The solutions were compared with each other and with the results of direct numerical simulations of the original nonlinear problem formulation, showing very good agreement for the waves of small amplitude, but also reasonably extending their relevance to the waves of moderate amplitude. In particular, the weakly-nonlinear solution has allowed us to generate and observe a stably propagating moderately-nonlinear longitudinal ``table-top" soliton. 

\section{Acknowledgements}

We would like to thank   O.E. Kurkina, A.V. Mikhailov, D.E. Pelinovsky, Y.A. Stepanyants and T.G. Talipova for useful references and discussions. We are also grateful to the organisers of the conference SCT-19 in honour of V.E. Zakharov's 80th Birthday where some of these discussions have taken place.
F.E.~Garbuzov and Y.M.~Beltukov acknowledge the financial support from the Russian Science Foundation under the grant no. 17-72-20201. 

\section*{Appendix A. Coefficients}
Here, the coefficients in the equation \eqref{eq_u0_fin} are given in terms of the Young modulus $E$ and the Poisson ratio~$\nu$ (instead of the Lame moduli $\lambda$ and $\mu$) and the third and fourth Landau moduli $A, B, C, D, E, F, G$:
\begin{align*}
%\label{alpha_1}
&\alpha_1 = \alpha_3 = \frac{1 + \nu}{4}, \quad \alpha_2 = -\frac{1 + \nu + \nu^2}{2}, \\
&\alpha_4 = \frac{(1 + \nu)^2}{48}, \quad \alpha_5 = -\frac{5 + 3\nu + 10\nu^3 - 4\nu^4 - 12 \nu^5}{48 (1 - \nu)}\\
&\alpha_6 = \frac{14 + 5\nu + 16\nu^2 - 8\nu^3 - 24\nu^4}{96(1 -\nu)}, \quad \alpha_7 = -\frac{6 + 13\nu + 14\nu^2 + 6\nu^3}{96 (1 + \nu)},\\
%\label{beta_1}
&\beta_1 = -\left(\frac32 E + A \left(1 - 2\nu^3\right) + 3B\left(1 + 2\nu^2\right) (1 - 2\nu) + C (1 - 2\nu)^3\right),\\
\begin{split}
&\beta_2 = 4(B + C)^2 - E\left(2A + 6B + 2C + 4(D+F+G+H) + \frac{E}{2}\right) \\
&\qquad +4\nu\left(-5B^2 - 14BC - 9C^2 + E(3B + 
3C + 2D + 4F + 8H)\right)\\
&\qquad +4\nu^2\left(18B^2 + 44BC + 30C^2 + 2A(B + C) - E(3B + 6C + 6F + 4G + 24H)\right)\\
&\qquad + 4\nu^3\left(- 32B^2 - 76BC - 40C^2 - 6AB - 10AC + E(A + 6B + 4C + 2D + 8F + 32H)\right)\\
&\qquad +4\nu^4\left(A^2 + 28 B^2 + 40BC + 12A(B + C) - 
4E(D + 2F + G + 4H)\right)\\
&\qquad - 4\nu^5\left(A^2 + 4 A (B - 2 C) - 4 (3 B^2 + 20 B C + 12 C^2)\right) - 8\nu^6(A + 6B + 4C)^2
\end{split}\\
\begin{split}
&\widetilde J_0 = \frac{1}{E}\big[ \gamma_1 \left(U_{0x} U_{0tt}\right)_{tt} + \gamma_2 \left(U_{0tt}^2\right)_x + \gamma_3 \left(U_{0x} U_{0tt}\right)_{xx} + \gamma_4 \left(U_{0x}^2\right)_{xtt} + \gamma_5 \left(U_{0x} U_{0xtt}\right)_x + \gamma_6 \left(U_{0x}^2\right)_{xxx} \\
&\qquad\quad + \gamma_7 \left(U_{0xx}^2\right)_x\big],
\end{split}\\
&\gamma_1 = -\frac{1+\nu}{4}\left[2E + A(1 - \nu^2) + 2B(1 + \nu)(1 - 2\nu)\right], \quad \gamma_2 = \frac{\gamma_1}{2}, \quad \gamma_3 = -4\gamma_1,\\
\begin{split}
&\gamma_4 = -\frac{1+\nu}{8}\big[E + A + 2B + 2\nu(B + 2C) - \nu^2(A + 20B + 24C - 2E) + 4\nu^3(A + 10B + 12C) \\
&\qquad\qquad- 8\nu^4(A + 6B + 4C)\big],
\end{split}\\
\begin{split}
&\gamma_5 = -\frac{1+\nu}{4}\big[5E + 3A + 10B + 4C - 2\nu(9B + 10C) + \nu^2(A + 12B + 24C + 2E) \\
&\qquad\qquad- 4\nu^3(A + 2B - 4C) - 8\nu^4(A + 6B + 4C)\big],
\end{split}\\
\begin{split}
&\gamma_6 = -\frac{1}{8}\big[4E + 3A + 10B + 4C + \nu(3A - 12B - 20C + 4E) + \nu^2(A + 6B + 24C + 2E) \\
&\qquad\qquad- \nu^3(7A + 20B - 16C) - 8\nu^4(A + 6B + 4C)\big],
\end{split}\\
\begin{split}
&\gamma_7 = -\frac{1}{8}\big[8E + 5A + 18B + 8C + \nu(A - 36B - 44C + 2E) + \nu^2(3A + 42B + 72C + 2E) \\
&\qquad\qquad- \nu^3(13A + 60B + 16C) - 8\nu^4(A + 6B + 4C)\big],
\end{split}
\end{align*}
The relations between the Landau and Murnaghan moduli can be found in (\ref{coefs}).

%\section*{Conflict of interests} The authors declare that they have no conflicts of interests.

%\section*{References}


\begin{thebibliography}{99}
\bibitem{W} G.B. Whitham, Linear and Nonlinear Waves, Wiley, New York, 1974.
\bibitem{AS} M.J. Ablowitz, H. Segur, Solitons and the Inverse Scattering Transform, SIAM, Philadelphia, 1981.
\bibitem{NMPZ} S. Novikov, S.V. Manakov, L.P. Pitaevskii, V.E. Zakharov, Theory of solitons: The Inverse Scattering Method, Springer, USA, 1984.
\bibitem{N} A.C. Newell, Solitons in Mathematics and Physics, SIAM, USA, 1985.
\bibitem{Maugin} G.A. Maugin, Nonlinear waves in elastic crystals, Oxford University Press, Oxford, 1999.
\bibitem{Dai} H.-H. Dai, Z. Cai, Phase transition in a slender cylinder composed of an incompressible elastic material. I. Asymptotic model equation, \textit{Proc. Roy. Soc. A} 462 (2006) 419-438.
\bibitem{M} A. Mayer, 
%Surface acoustic waves in nonlinear elastic media, \textit{Phys. Reports} 256 (1995) 257 - ?.
Nonlinear surface acoustic waves: Theory, \textit{Ultrasonics} 48 (2008) 478-481.
\bibitem{HL}
P. Hess, A.M. Lomonosov, Solitary surface acoustic waves and bulk solitons in nanosecond and picosecond laser ultrasonics, \textit{Ultrasonics} 50 (2010) 167-171.
\bibitem{E1} J. Engelbrecht, A. Salupere and K. Tamm, Waves in microstructured solids and the Boussinesq paradigm, \textit{Wave Motion} 48 (2011) 717-726.
\bibitem{P}
A. Pau, F. Lanza di Scalea, Nonlinear guided wave propagation in prestressed plates, \textit{J. Acoust. Soc. Am.} 137 (2015) 1529-1540.
\bibitem{E2} T. Peets, K. Tamm, J. Engelbrecht, On the role of nonlinearities in the Boussinesq-type wave equations, \textit{Wave Motion} 71 (2017) 113-119.
\bibitem{NS}
G.A. Nariboli, A. Sedov, Burgers-Korteweg de Vries equation for viscoelastic rods and plates, \textit{J. Math. Anal. Appl.} 32(3) (1970) 661-677.
\bibitem{OS}
L.A. Ostrovsky, A.M. Sutin, Nonlinear elastic waves in rods, \textit{PMM} 41 (1977) 531-537.
\bibitem{S1}
A.M. Samsonov, Structural optimization in nonlinear wave propagation problems. In: \textit{Structural Optimization under Dynamical Loading. Seminar and Workshop for Junior Scientists}, U. Lepik ed., Tartu University Press, 75-76 (1982).
\bibitem{S2}
A.M. Samsonov, Soliton evolution in a rod with variable cross section, \textit{Sov. Physics - Doklady} 29 (1984) 586-587.
\bibitem{SP}
A.M. Samsonov, A.V. Porubov, Refinement of the model for the propagation of longitudinal strain waves in a rod with nonlinear elasticity, \textit{Tech. Phys. Lett.} 19(6) (1993) 365-366.
\bibitem{PV}
A.V. Porubov, M.G. Velarde, Dispersive - dissipative solitons in nonllinear solids, \textit{Wave Motion} 31(3) (2000) 197-207.
%\bibitem{E_book}
%V.I. Erofeev, V.V. Kazhaev, N.P. Semerikova, \textit{Waves in rods: dispersion, dissipation, nonlinearity}, Fizmatlit, Moscow, 2002 (in Russian).
\bibitem{S_book} A.M. Samsonov, Strain solitons in solids and how to construct them, Chapman \& Hall/CRC, Boca Raton, 2001.
\bibitem{P_book} A.V. Porubov, Amplification of nonlinear strain waves in solids, World Scientific, Singapore, 2003.
\bibitem{Murnaghan}
F.D. Murnaghan, Finite deformation of an elastic solid, John Wiley and Sons, New York, 1951.
%\bibitem{BBM} T.B. Benjamin, J.L. Bona, and J.J. Mahony, Model equations for long waves in %nonlinear dispersive systems, \textit{Philos. Trans. R. Soc. London, Ser. A} 272(1220) (1972) 47-78.
%\bibitem{DC} H.-H. Dai, and Z. Cai, Uniform asymptotic analysis for transient waves in a pre-stressed %compressible hyperelastic rod, \textit{Acta Mechanica} 139 (2000) 201-230.
\bibitem{DF} H.-H. Dai, X. Fan, Asymptotically approximate model equations for weakly nonlinear long waves in compressible elastic rods and their comparisons with other simplified model equations, \textit{Maths. Mechs. Solids} 9 (2004) 61-79.
\bibitem{GKS} F.E. Garbuzov, K.R. Khusnutdinova, I.V. Semenova, On Boussinesq-type models for long longitudinal waves in elastic rods, \textit{Wave Motion} 88 (2019) 129-143. 
\bibitem{KSZ}
K.R. Khusnutdinova, A.M. Samsonov, A.S. Zakharov, Nonlinear layered lattice model and generalized solitary waves in imperfectly bonded structures, \textit{Phys. Rev. E} 79(5) (2009) 056606.
\bibitem{KS}
K.R. Khusnutdinova, A.M. Samsonov, Fission of a longitudinal strain solitary wave in a delaminated bar, \textit{Phys. Rev. E} 77 (2008) 066603.
\bibitem{KT1}
K.R. Khusnutdinova, M.R. Tranter, Modelling of nonlinear wave scattering in a delaminated elastic bar, \textit{Proc. R. Soc. A} 471 (2015) 20150584.
\bibitem{KT2}
K.R. Khusnutdinova, M.R. Tranter, On radiating solitary waves in bi-layers with delamination and coupled Ostrovsky equations, \textit{Chaos} 27 (2017) 013112.
\bibitem{JAP2010}
G.V. Dreiden, K.R. Khusnutdinova, A.M. Samsonov, and I.V. Semenova, Splitting induced generation of soliton trains in layered waveguides, \textit{J. Appl. Phys.} 107 (2010) 034909.
\bibitem{JAP2012}
G.V. Dreiden, K.R. Khusnutdinova, A.M. Samsonov, and I.V. Semenova, Bulk strain solitary waves in bonded layered polymeric bars with delamination, \textit{J. Appl. Phys.} 112 (2012) 063516.
\bibitem{Mathematica}
MATHEMATICA and WOLFRAM MATHEMATICA are registered trademarks of Wolfram Research Inc. (www.wolfram.com)
\bibitem {K} Y. Kodama, Normal forms for weakly dispersive wave equations, \textit{Phys. Lett. A} 112 (1985) 193-196.
\bibitem{FL} A.S. Fokas, Q.M. Liu, Asymptotic integrability of water waves, \textit{Phys. Rev. Lett.} 77 (1996) 2347-2351.
\bibitem{HK} Y. Hiraoka and Y. Kodama, Normal forms and solitons, In A.V. Mikhailov ed. \textit{Integrability}, Lecture Notes in Physics, Springer-Verlag, Berlin, Heidelberg, 2009.
\bibitem{LL} L.D. Landau and E.M. Lifshitz, Theory of Elasticity, Pergamon Press, Oxford, 1986.
\bibitem{CLS} P.A. Clarkson, R.J. LeVeque, R. Saxton, Solitary wave interaction in elastic rods, \textit{Stud. Appl. Math.} 75 (1986) 95-122.
\bibitem{SCL1} M.P. Soerensen, P.L. Christiansen, P.S. Lomdahl, Solitary waves in non-linear elastic rods, I, \textit{I. J. Acoust. Soc. Amer.} 76 (1984) 871-879.
\bibitem{SCL2} M.P. Soerensen, P.L. Christiansen, P.S. Lomdahl, Solitary waves in non-linear elastic rods, II, \textit{I. J. Acoust. Soc. Amer.} 81 (1984) 1718-1722.
\bibitem{PM} A.V. Porubov, G.A. Maugin, Longitudinal strain solitary waves in presence of cubic non-linearity, \textit{Intern. J. Non-Lin. Mech.} 40 (2005) 1041-1048.
\bibitem{B} D.J. Benney, Long non-linear waves in fluid flows, \textit{J. Math. Phys.} 45 (1966) 52-63.
\bibitem{LB} C.-Y. Lee and R.C. Beardsley, The generation of long nonlinear internal waves in weakly stratified shear flows, \textit{J. Geophys. Res.} 79 (1974) 453-457.
\bibitem{KB} C. Koop and G. Butler, An investigation of internal solitary waves in a two-fluid system, \textit{J. Fluid Mech.} 112 (1981) 225-251.
\bibitem{LY} K. Lamb and L. Yan, The evolution of internal wave undular bores: comparisons of a fully nonlinear numerical model with weakly nonlinear theory, \textit{J. Phys. Oceanography} 26 (1996) 2712-2734.
\bibitem{MS} T.R. Marchant and N.F. Smyth, The extended Korteweg - de Vries equation and the resonant flow of a fluid over topography, \textit{J. Fluid Mech.} 221 (1990) 263-288.
\bibitem{GPP} R. Grimshaw, E. Pelinovsky, and O. Poloukhina, Higher-order Korteweg - de Vries models for internal solitary waves in a stratified shear flow with a free surface, \textit{Nonlinear Processes Geophys.} 9 (2002) 221-235.
\bibitem{GKKS} A.R. Giniatullin, A.A. Kurkin, O.E. Kurkina, and Y.A. Stepanyants, Generalised Korteweg - de Vries equation for internal waves in two-layer fluid, \textit{Fundum. Appl. Hydrophysics} 7 (2014) 16-28.
\bibitem{KRI} A. Karczewska, P. Rozmej, and E. Infeld, Shallow-water soliton dynamics beyond the Korteweg - de Vries equation, \textit{Phys. Rev. E} 90 (2014) 012907.
\bibitem{KST} K.R. Khusnutdinova, Y.A. Stepanyants, M.R. Tranter, Soliton solutions to the fifth-order Kortweg - de Vries equation and their applications to surface and internal water waves, \textit{ Phys. Fluids} 30 (2018) 022104.
\bibitem{D} H.-H. Dai, Model equations for non-linear dispersive waves in a compressible Mooney-Rivlin rod, \textit{Acta mech.} 127 (1998) 193-207.
\bibitem{DH} H.-H. Dai, Yi Huo, Solitary shock waves and other travelling waves in a general compressible hyperelastic rod, \textit{Proc. R. Soc. Lond. A} 456 (2000) 331-363.
\bibitem{MS2} T.R. Marchant and N.F. Smyth, Soliton interaction for the extended Korteweg - de Vries equation, \textit{J. Appl. Math.} 56 (1996) 157-176.
\bibitem{MS3} T.R. Marchant and N.F. Smyth, An undular bore solution for the higher-order Korteweg - de Vries equation, \textit{J. Phys. A: Math. Gen.} 39 (2006) L563-L569.
\bibitem{PS} O.E. Polouknina, A.V. Slyunyaev, Corrected evolution model on the basis of Gardner equation for internal waves in a stratified fluid, \textit{Izvestiya A.M. Prokhorov AIN, Appl. Maths and Mechs.} 18 (2006) 82 - 91 (in Russian).
\bibitem{RKK_book} E.A. Ruvinskaya, O.E. Kurkina, A.A. Kurkin, Dynamics of nonlinear internal gravitational waves in layered fluids, State Technical University of Nizhni Novgorod, 2014 (in Russian).
\bibitem{PPL} E.N. Pelinovskii, O. Polukhina, and K. G. Lamb, \textit{Oceanology} (Engl. Transl.) 40 (2000) 757-766.
% Пелиновский Е.Н., Полухина О.Е., Лэмб К. Нелинейные внутренние волны в океане, стратифицированном по плотности и течению. // Океанология. 2000. Т. 40. №6. С. 805-815.
\bibitem{L_et_al} K.G. Lamb, O. Polukhina, T. Talipova, E. Pelinovsky, W. Xiao, A. Kurkin, Breather generation in fully nonlinear models of a stratified fluid, \textit{Phys. Rev. E} 75 (2007) 046306.
\bibitem{M_et_al} V. Maderich, T. Talipova, R. Grimshaw, E. Pelinovsky, B.H. Choi, I. Brovchenko, K. Terletska, and D.C. Kim, The transformation of an interfacial solitary wave of elevation at a bottom step, \textit{ Nonlin. Processes Geophys}. 16 (2009) 33-42.
\bibitem{M_et_al1} V. Maderich, T. Talipova, R. Grimshaw, K. Terletska, I. Brovchenko, E. Pelinovsky and B.H. Choi, Interaction of a large amplitude interfacial solitary wave of depression with a bottom step, \textit{ Phys. Fluids} 22 (2010) 076602.
\bibitem{T_et_al} T. Talipova, K. Terletska, V. Maderich, I. Brovchenko, K. T. Jung et al., Internal solitary wave transformation over a bottom step: Loss of energy, \textit{ Phys. Fluids} 25 (2013) 032110.
\bibitem{CHQZ3} C. Canuto, M.Y. Hussaini, A. Quarteroni, T.A. Zang, Spectral Methods. Evolution to Complex Geomenties and Applications to Fluid Dynamics, Springer-Verlag, Berlin, 2007.
\end{thebibliography}
\end{document}